\documentclass[usenatbib,usegraphicx]{mn2e}

\usepackage{subfigure}
\usepackage{longtable}
\usepackage{graphicx}
\usepackage{graphics}
\usepackage{keyval}
\usepackage{trig}
\usepackage[american]{babel}
\usepackage{url}
\usepackage{color}
\usepackage{amsmath}
\usepackage{bm}

\def\beq{\begin{equation}}
\def\eeq{\end{equation}}
\def\bey{\begin{eqnarray}}
\def\eey{\end{eqnarray}}

\def\kpc{\, {\rm kpc} }
\def\msun{M_\odot}

\def\lsim{\mathrel{\raise.3ex\hbox{$<$\kern-.75em\lower1ex\hbox{$\sim$}}}}
\def\gsim{\mathrel{\raise.3ex\hbox{$  $\kern-.75em\lower1ex\hbox{$\sim$}}}}
\def\Msun{M_\odot}

\def\tz{t_\mathrm{0}}
\def\te{t_\mathrm{E}}
\def\uz{u_\mathrm{0}}

\def\thetae{\theta_\mathrm{E}}
\def\fs{F_\mathrm{S}}
\def\fb{F_\mathrm{B}}
\def\ds{D_\mathrm{S}}
\def\dl{D_\mathrm{L}}
\def\tentry{t_\mathrm{entry}}
\def\texit{t_\mathrm{exit}}
\def\sentry{s_\mathrm{entry}}
\def\sexit{s_\mathrm{exit}}
\def\a0{A_{\mathrm{0}}}

\def\ai{A_{\mathrm{I}}}

\newcommand\Eq[1]{Eq.~(\ref{#1})}
\newcommand\Fig[1]{Fig.~\ref{#1}}
\newcommand\Tab[1]{Table~\ref{#1}}
\newcommand\Sec[1]{Sec.~\ref{#1}}

\def\tE {t_\mathrm{E}}

\def\to {t_0}

\def\rhoS {\rho_\mathrm{*}}

\title{A systematic fitting scheme for caustic-crossing microlensing events}
\author[N. Kains et al.]
{N.Kains$^{1,2,}$\thanks{email:nk87@st-andrews.ac.uk}, 
A. Cassan$^{1,3}$, 
K.Horne$^{1,2}$, 
M.D. Albrow$^{1,4}$,
S. Dieters$^{1,5}$,  \and
P. Fouqu\'{e}$^{1,6}$, 
J. Greenhill$^{1,7}$,
A. Udalski$^{8,9}$, 
M. Zub$^{1,3}$\thanks{Member of International Max Planck Research School for Astronomy and Cosmic Physics at the University of Heidelberg}, 
D.P. Bennett$^{1,10}$,\and
M. Dominik$^{2}$\thanks{Royal Society University Research Fellow}, 
J. Donatowicz$^{1,11}$,
D. Kubas$^{1,12}$,   
Y. Tsapras$^{1,13}$,  
T. Anguita$^3$, \and
V. Batista$^{1,5}$,  
J.-P. Beaulieu$^{1,5}$, 
S. Brillant$^{1,12}$, 
M. Bode$^{1,13}$, 
D.M. Bramich$^{1,14}$,  \and
M. Burgdorf$^{1,13}$, 
J.A.R. Caldwell$^{1,15}$,
K.H. Cook$^{1,16}$, 
Ch. Coutures$^{1,17}$,   \and
D. Dominis Prester$^{1,18}$, 
U.G. J\o rgensen$^{1,19}$,  
S. Kane$^{1,20}$,  
J.B. Marquette$^{1,5}$, \and
R. Martin$^{1,21}$, 
J. Menzies$^{1,22}$, 
K.R. Pollard$^{1,4}$,
N. Rattenbury$^{1,23}$,  
K.C. Sahu$^{1,24}$,  \and
C. Snodgrass$^{1,12}$, 
I. Steele$^{1,11}$,  
C. Vinter$^{1,19}$, 
J. Wambsganss$^{1,3}$, 
A. Williams$^{1,21}$,  \and
M. Kubiak$^{8,9}$, 
G. Pietrzy{\'n}ski$^{8,9,25}$,
I. Soszy{\'n}ski$^{8,9}$, 
O. Szewczyk$^{8,9,25}$,   \and
M.K. Szyma{\'n}ski$^{8,9}$,  
K. Ulaczyk$^{8,9}$ 
{\L}.Wyrzykowski$^{13,26}$ \\ \\
$^{1}$PLANET/RoboNet collaborations\\
$^{2}$SUPA, School of Physics and Astronomy, University of St. Andrews, North Haugh, St Andrews, KY16 9SS, United Kingdom\\
$^{3}$Astronomisches Rechen-Institut (ARI), Zentrum f\"{u}r Astronomie (ZAH), Heidelberg University,\\  M\"{o}nchhofstra\ss e 12-14, 69120 Heidelberg, Germany\\
$^{4}$University of Canterbury, Department of Physics and Astronomy, Private Bag 4800, Christchurch, New Zealand\\
$^{5}$Institut d'Astrophysique de Paris, UMR7095 CNRS, Universit\'{e} Pierre \& Marie Curie, 98bis Boulevard Arago, 75014 Paris, France\\
$^{6}$LATT, Universit\'{e} de Toulouse, CNRS, 14 avenue Edouard Belin, F-31400 Toulouse, France\\
$^{7}$School of Mathematics and Physics, University of Tasmania, Private Bag 37, Hobart, Tasmania 7001, Australia\\
}

\begin{document}

\date{Accepted ... Received ... ; in original form ...}

\pagerange{\pageref{firstpage}--\pageref{lastpage}} \pubyear{2008}

\maketitle

\label{firstpage}
\begin{abstract}
We outline a method for fitting binary-lens caustic-crossing
  microlensing events based on the alternative model parameterisation
  proposed and detailed in Cassan~(2008). As an illustration of our methodology, we
  present an analysis of OGLE-2007-BLG-472, a double-peaked Galactic
  microlensing event with a 
source crossing the whole caustic structure in less than three days. In order to identify all
possible models we conduct an extensive search of the parameter space, followed by a refinement of the parameters with a Markov Chain-Monte
Carlo algorithm. We find a number of low-$\chi^2$ regions in the parameter space, which lead to several
distinct competitive best models. We examine the parameters for each
of them, and estimate their physical properties.
We find that our fitting strategy locates several minima that are difficult to find with other modelling strategies and is therefore a more appropriate method to fit this type of events.

\end{abstract}


\section{Introduction}\label{sec:intro}

  Gravitational microlensing \citep{paczynski86} occurs when the light from a
  source star is deflected by a massive compact object between the
  source and the observer, leading to an apparent brightening of the
  source, typically lasting a few days to a few weeks. When the
  deflecting body has multiple components, such as a planet orbiting its
  host star, there can be perturbations to the brightening pattern of
  observed sources. These perturbations can be large even when caused by
  low-mass objects, making them detectable using small ground-based
  telescopes. Modelling these lightcurve anomalies can lead to the
  detection of subtle effects,  allowing for measurements of properties such as the source star limb-darkening
  coefficients (e.g. \citealt{cassan04-ob04069}),
  the mass of stars with no visible
  companions (e.g. \citealt{ghosh04}), and the detection of extrasolar
  planets, as suggested by \cite{maopaczynski91} and first
  achieved in 2003 \citep{bond04-ob235}.

  Nevertheless, anomalous microlensing events usually require
  very detailed analysis for a full characterisation of their nature to be possible, making them challenging.
  This applies in particular to a class of microlensing events which
  display caustic crossing features in their lightcurves. 
  These events are of primary interest, because they
  account for around ten percent of the overall number of detected microlenses,
  and they represent an important source of information on physical properties of binary stars
 \citep{jaroszynski06}. 
  However there exist several
  degeneracies that affect the modelling of this type of events: without a robust modelling
  scheme and a full exploration of the parameter space, it is impossible to pin down the true nature of a given event.
  In addition to this, calculations of anomalous microlensing models for extended sources are very demanding computationally.

  Given these issues, brute force is not an option when modelling
  caustic-crossing events, and one has to devise ways of speeding up
  calculations, for example by excluding regions of parameter space
  which cannot reproduce features that appear in data sets. 
  A way to achieve this is to use a non-standard parameterisation
  of the binary-lens models that ties them directly to data features,
  as proposed by \cite{cassan08cf}, which we recall below.

  In this paper, we present our method for
  exploring the parameter space, and describe our approach to find all
  possible models for a given event (Sec. \ref{sec:modelling}). We then
  use OGLE-2007-BLG-472, a microlensing event observed in
  2007 by the OGLE and PLANET collaborations, as an illustration of our
  methodology applied on a binary lens event which intrinsically harbors 
  many ambiguities (Sec. \ref{sec:event}). 
  We finally discuss the implications of the
  individual competitive models that we find in order to discriminate between
  realistic microlensing scenarii.

\section{Binary-lens events fitting scheme}\label{sec:modelling}

\subsection{Parameterisation of binary lens lightcurves}

  A static binary lens is usually described by the mass ratio $q<1$ of
  the two lens components and by their separation $d$,
  expressed in units of the angular Einstein 
  radius \citep{einstein36}, 
  \begin{equation}
    \label{eq:thetae}
    \thetae=\sqrt{\frac{4GM}{c^2}\left(\frac{D_\mathrm{S}-D_\mathrm{L}}{D_\mathrm{S}D_\mathrm{L}}\right)}\, ,
  \end{equation}
  where $M$ is the mass of the lens, and $\dl$ and $\ds$ are the distances
  to the lens and the source respectively.
  Such a lens produces caustics where the magnification of the 
  source diverges to infinity for a perfect point source. The
  positions, sizes and shapes of the caustics depend on $d$ and $q$.
  For the binary lens case, caustics can exist in three different
  topologies, usually referred as \textit{close}, \textit{intermediate}
  and \textit{wide}; bifurcation values between these topologies are
  analytical expressions relating $d$
  with $q$ \citep{erdlschneider93}. 
  In the close regime, there are three caustics: a \textit{central}
  caustic near the primary lens 
  component, and two \textit{secondary} caustics which lie off the
  axis passing through both lens 
  components. In the intermediate case, there is only one large
  caustic on the axis. In the wide case, 
  there is a \textit{central} as well as a \textit{secondary} caustic,
  both on the axis. The limits between these configurations 
  are indicated as the dashed lines in e.g \Fig{fig:chi2mapC}
  (see also Fig.~1 of \cite{cassan08cf}). 
  
  The description of the lightcurve itself requires four more
  geometrical parameters in addition to $d$ and $q$. In the current standard parameterisation of binary lens lightcurves, these are the source trajectory's
  angle $\alpha$ with the axis of symmetry of the lens,
  the time of closest source-lens approach to the binary lens
  centre-of-mass $\tz$, the 
  Einstein radius crossing time $\te$ and the source-lens separation at
  closest approach $\uz$ (in units of $\thetae$). Finally for a
  uniformly bright finite size source star, we add a further
  parameter, the source size $\rhoS$ in units of $\thetae$.
  However, and as discussed in \cite{cassan08cf}, this
  parameterisation is not well adapted to conducting a full search of the
  parameter space, because the value of the parameters cannot be directly
  related to features present in the lightcurve, namely
  caustic crossings for the type of events we are discussing in this paper. Consequently, most of the probed models in a
  given fitting process do not exhibit the most obvious features in
  the lightcurve, leading to very inefficient modelling.

  To avoid this drawback, \cite{cassan08cf} introduced a new
  parameterisation in place of $\alpha, \tz, \uz$ and $\te$ which is
  closely related to the appearance of caustic crossing features in the
  lightcurve. The caustic entry is then defined by a date $\tentry$ 
  when the source center crosses the caustic\footnote{Alternatively,
    any other point at a fixed position from the source center can be
    defined as a reference.} and its corresponding (two-dimensional)
  coordinate $\zeta_{\rm entry}$ on the source plane. 
  However, since by definition this point is
  located on a caustic line, \cite{cassan08cf} introduced a (one-dimensional)
  \textit{curvilinear abscissa} $s$ which locates the crossing point
  directly on the caustic, so that $\zeta_{\rm entry} \equiv \zeta(\sentry)$.
  A given caustic structure 
  is fully parameterised by $0 \leq s \leq 2$. The caustic entry
  is then characterised by a pair of parameters $(\tentry, \sentry)$,
  and in the same way the caustic exit by $(\texit, \sexit)$. These
  four parameters (in addition to $d, q$ and $\rhoS$) which describe the caustic
  crossings therefore also define an alternative
  parameterisation of the binary lens, far better fitted to
  describing the problem at hand.

\subsection{Exploration of the parameter space}\label{sec:exploration}

  We start by exploring a wide region of the parameter space 
  with a $(d,q)$ grid regularly
  sampled in logarithmic scale. This choice comes from the fact that the
  size of the caustic structures behave like power-laws of the lens
  separation and mass ratio, and so do the corresponding lightcurve
  anomalies. We fit for the remaining model parameters $\tentry$,
  $\texit$, $\sentry$, $\sexit$ and $\rhoS$, $(d,q)$ being held
  fixed. From this, we then build a $\chi^2(d,q)$ map that we use to
  locate the best-fit $(d,q)$-regions.
  As mentioned previously, there exist binary lens
  configurations which involve central and secondary caustics. In these
  cases (\textit{i.e.} in the wide and close
  binary cases) and following \cite{cassan08cf}, we study
  separately models where the source crosses
  the central or the secondary caustic by building two $\chi^2(d, q)$ maps, corresponding to
  each configuration. 

  In order to sample efficiently and extensively $\sentry$ and
  $\sexit$ (which determine the source 
  trajectory), we use a genetic algorithm \citep[e.g.][]{pikaia} 
  that always keeps the best model from one generation to the next one (\textit{elitism}). In
  fact, since we consider only models displaying caustics at the right
  positions, there are a couple of local minima 
  associated with different $(\sentry, \sexit)$ pairs which would usually be missed by other minimisation methods, while a genetic
  algorithm naturally solves this problem in an efficient
  way. However, since such an algorithm never converges exactly to the best model,
  we finally refine the model by performing a Markov-Chain Monte-Carlo
  (hereafter MCMC) fit: we start several chains and use the criterion by
  \cite{geweke92} to assess convergence to a stationary posterior 
  distribution of the parameter probability densities.
  
  From the obtained $\chi^2$ maps, we then identify all the local
  minima regions and use the corresponding best models found on the
  $(d,q)$ grid as starting points to refine the parameters, including
  $(d,q)$ that we now allow to vary. Since the fit is performed within a
  minimum $\chi^2$ region, the fitting process is very stable and
  fast.

\section{Application to OGLE-2007-BLG-472}\label{sec:event}

\subsection{Alert and photometric follow-up}

On 19 August 2007, the OGLE Early Warning System \citep{udalski03ews} flagged microlensing candidate event OGLE-2007-BLG-472 at right ascension $\alpha_{2000.0}=$ 17:57:04.34, and declination $\delta_{2000.0}=$ -28:22:02.1 or $l=1.77^{\circ}$, $b=-1.87^{\circ}$.

\begin{figure}
  \centering
  \includegraphics[width=6cm, angle=270]{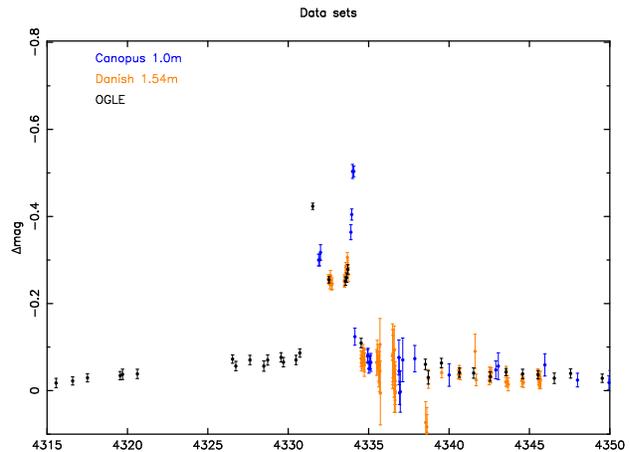}
  \caption{OGLE, UTas and Danish data set for OGLE-2007-BLG-472 data sets. Data points are plotted with 1-$\sigma$ error bars. \label{fig:pspl-noblend}. The $x-$axis is time in HJD-2450000.}
\end{figure}

The OGLE lightcurve has an instrumental baseline magnitude I=16.00, which may differ from the calibrated magnitude by as much as 0.5 magnitudes. Lensing by the star in the point source-point lens (hereafter PSPL) approximation
accounts for a broad rise and fall in the lightcurve, peaking around MHJD\footnote{MHJD=HJD-2450000}=4334.0
with an apparent half-width at half-peak of about 10 days (\Fig{fig:pspl-noblend}).  Although the observed OGLE flux rises only by 0.06 mag in the non-anomalous part of the lightcurve, the shape of the curve hints that blending is important for this target, with only $\sim$ 12\% of the baseline flux due to the un-magnified source. 

On 19 August (MHJD=4331.5) an OGLE data point showed sudden brightening of the source, with subsequent PLANET (UTas Mt. Canopus 1.0m telescope in Tasmania and Danish 1.54m telescope at La Silla, Chile) and OGLE data indicating what appears to be a fold 
caustic crossing by the source, ending with a PLANET UTas data point on August 21 (MHJD=4334.1). The caustic entry is observed by a single OGLE point, while the caustic exit is well covered by our UTas data set (\Fig{fig:pspl-noblend}). Treating the lightcurve as the addition of an anomaly to a PSPL lightcurve, the underlying PSPL curve then apparently reaches peak magnification on August 22 (MHJD=4335.45). Particularly crucial in our data set is the UTas observation taken within a few hours of the caustic exit, which puts constraints on the position of the caustic exit on the lightcurve, and on the size of the source. Although some V-band observations were taken, the V lightcurve of this event is flat and does not allow us to place constraints on the V flux parameters.

\begin{table}
  \begin{center}
    \begin{tabular}{ccc}
      \hline
      Telescope & Data & Error bar rescaling factor \\
      \hline
      UTas 1.0m	    & 34	& 1.79	\\
      Danish 1.54m  & 84	& 1.55	\\
      OGLE          & 857	& 1.21	\\ \hline 
    \end{tabular}
    \caption{Datasets and error bar rescaling factors.}
    \label{datatable}
  \end{center}
\end{table}

\subsection{Data reduction}

We reduced the PLANET data for this event using the data reduction pipeline pysis3.0 (Albrow 2008). This pipeline uses a kernel as a discrete pixel array, as proposed by \cite{bramich08kernel}, rather than a linear combination of basis functions.This has the advantage that it removes the need for the user to select basis functions manually, which can lead to problems if inappropriate functions are chosen. In addition to this, the pixel array kernel copes better with images that are not optimally aligned. The result of using this pipeline is a better reduction than was obtained with other methods. We kept all points with seeing $<$3.5 arcseconds. Although some dubious points remain with this simple cut, the size of their associated error bars reflects their lack of certainty and ensures their weight in modelling procedures is appropriately reduced. Our final data set consists of 34 UTas data points, 84 points from the Danish 1.54m telescope, and 857 points from OGLE (Table \ref{datatable}).

\subsection{Modelling OGLE-2007-BLG-472}

\begin{figure*}
  \centering
  \includegraphics[width=12cm]{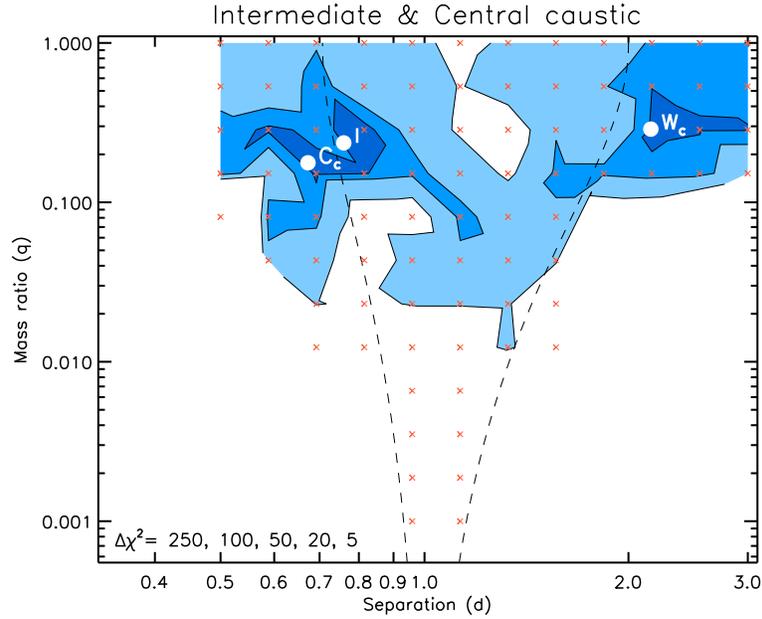}
  \caption{
    $\chi^2(d,q)$ map for the intermediate and
    central caustic configurations. Contour lines and minima regions
    (in blue shades) are plotted at $\Delta\chi^2 = 5, 20, 50, 100, 250$. The two dashed curves
    are the separation between the close, intermediate and wide regimes. The models are labelled
    and marked with white filled circles. }
  \label{fig:chi2mapC}
\end{figure*}

\begin{figure*}
  \centering
  \includegraphics[width=12cm]{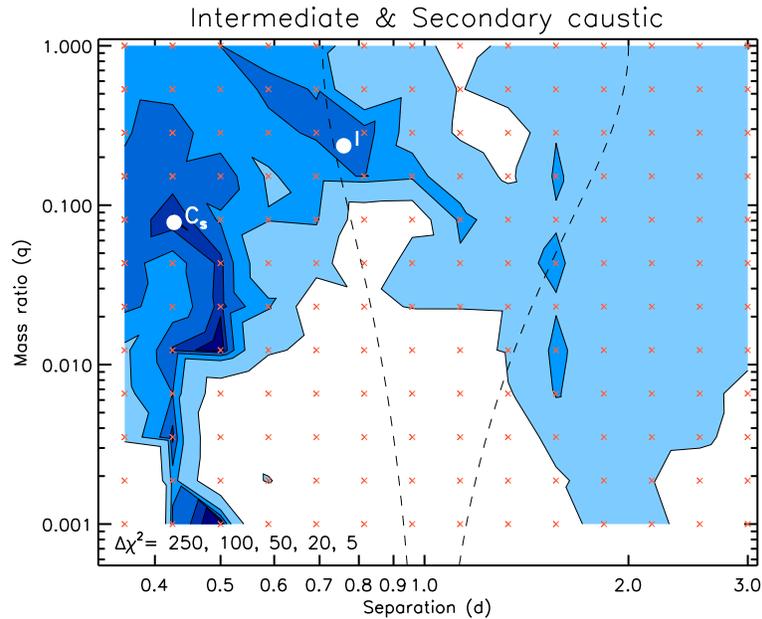}
  \caption{
    Same as \Fig{fig:chi2mapC} for the the intermediate and secondary caustic
    configuration. }
  \label{fig:chi2mapS}
\end{figure*}

\begin{figure*}
  \centering
  \includegraphics[width=13cm]{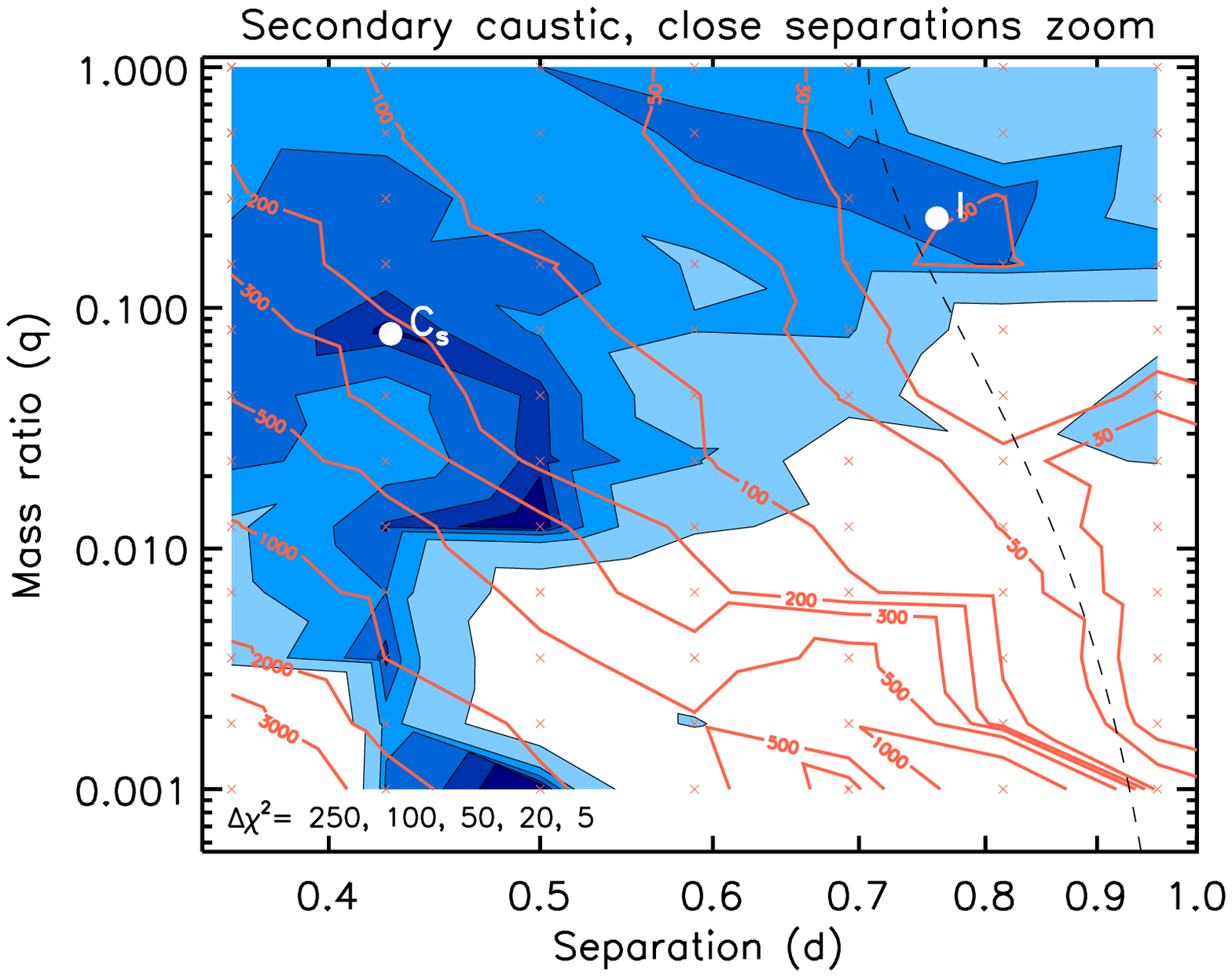}
  \caption{
    Map of the value of $\te$ in the $(d,q)$ plane for converged models at each
    grid point, superimposed on the $\chi^2$ map, zoomed in on the close regime part of parameter
    space. Contours lines (orange) are labeled with their corresponding value of $\te$ while $\chi^2$
    contour lines are plotted at $\Delta\chi^2 =5$, $20$, $50$, $100$, $250$ and filled with
    gradual shades of blue. The dashed curve is the separation between the close and intermediate
    regimes. The models of \Tab{tab:partable} are labelled and marked with white filled circles.} 
  \label{fig:tEmap}
\end{figure*}

After a first exploration of the parameter space, we find a best model (close to model $C_c$, see below) which we
use as a basis to rescale our error bars. In fact, these can vary rather widely from one telescope
to another and are often underestimated by photometry software. Ignoring this effect would
misrepresent the relative importance of the data sets. 
From this step, we choose the rescaling factors shown
in \Tab{datatable}, obtained by setting $\chi^2/\mathrm{d.o.f.} \simeq 1$ for each data set. We then
use the rescaled data to perform a new parameter space exploration.

We then apply the fitting scheme detailed in section
\ref{sec:modelling} to our data sets. 
  In particular, we choose a spacing between
  the $(d,q)$ grid points of 0.070 in $\mathrm{log}\, d$ and 0.275 in
  $\mathrm{log}\, q$. For the genetic algorithm fit, we use a model
  population of 200 individuals evolving during 40 generations, which
  has proven to be enough to safely locate the regions of minimum $\chi^2$.
  Finite source effects are computed using the adaptive
  contouring method of \cite{dominik07ac}. 

The final $\chi^2(d,q)$ maps that we
obtain are plotted in \Fig{fig:chi2mapC} for the intermediate and 
central caustic configurations, and \Fig{fig:chi2mapS} for the intermediate and secondary
caustic. The red crosses show the underlying $(d,q)$ grid, and the blue shaded contours indicate values
of $\Delta\chi^2 =5$, $20$, $50$, $100$, $250$, where the reference model is $C_s$, the global
best-fitted model (as obtained in Section \ref{sec:bestfit}).

\subsection{Excluding minima}\label{sec:excluding}

\Fig{fig:tEmap} shows a zoom of this region (secondary caustic and close
configuration $d<1$), with an overplot of $\tE$ isocontours (orange lines) roughly equally spaced on a
logarithmic scale. With this fitting approach, we put no initial constraints on the Einstein time $\tE$, though it will
always remain physical ($\tE>0$). Since we are not using 
any Bayesian prior for this parameter, we find that very good fits to the
data are obtained with values of $\tE>300$ days, which correspond to the minimum region in the left lower
part  of \Fig{fig:chi2mapS}.
Such long Einstein times are not likely to happen commonly, and it may happen that
some of the values found for $\tz$ correspond to a lightcurve that reaches its peak well in the future; these
are very unlikely to be acceptable solutions. Hence, instead of using a {\it prior} for $\tE$ in the fitting
process, we adopt the {\it posterior} distribution of \cite{dominik06stoch}, from which we see that
$\tE>400$ days, well in the tail of the distribution, can be used as a cut for a model to be physically
plausible. Thus in the following, we will not consider solutions with values of 
$\tE$ greater than 400 days. This means that we will not include the low-$q$ ($q \sim 0.001$) minima in the following discussion.

Although a very well-covered lightcurve generally enables a good
characterisation of the deviation caused by the caustic approach or
crossing, degeneracies make finding a unique best-fitting model
difficult. In particular, \cite{griestsafizadeh98} and
\cite{dominik99b} identified a two-fold degeneracy in the projected
lens components separation parameter $d$, under the change $d
\leftrightarrow 1/d$, when $q \ll 1$. Moreover, \cite{kubas05} showed
that very similar lightcurves could arise for a source crossing the
secondary caustic of a wide binary system and for the central caustic
of a close binary system. These degeneracies cause widely separated
$\chi^2$ minima in the parameter space, which must then be located by
exploring the parameter space thoroughly. In addition to these
degeneracies, imperfect sampling can increase the number of local
$\chi^2$ minima; short event in particular are prone to
under-sampling, leading to difficulties in
modelling. OGLE-2007-BLG-472 is no exception, as shown in the next
section.

\subsection{Refining local minima} \label{sec:bestfit}

We see from \Fig{fig:chi2mapC} (intermediate and central caustic) that there are three broad local
minima in the region around the white filled circles marked as $C_c$, $I$ and $W_c$ (``I'', ``C'' and
``W'' for intermediate, close and wide models respectively, and subscript ``c'' for central caustic). In 
\Fig{fig:chi2mapS} (intermediate and secondary caustic), a best-fit region can easily be located
around the region marked $C_s$ (subscript ``s'' for secondary caustic), besides region $I$. 

\begin{table*}
  \caption{Best-fitting binary lens model parameters. The blending factor $g(I)=\fb(I)/\fs(I)$ is given for the OGLE data ($I$-band). 
  The error bars were rescaled for each telescope by the factor given in \Tab{datatable}, which lead to the rescaled $\chi^2$ indicated here. Physical parameters are also given for each model, for the case of a lens in the disk, and a lens in the bulge. These were calculated using the procedure detailed in \Sec{sec:lensprop}}
  \begin{tabular}{ccccccc}
    \hline
    Parameter 				& Model $C_s$ 					& Model $C_c$ 		& Model $I$				& Model $W_c$				&Units \\
    \hline
$\chi^2$ (rescaled $\sigma$)	& $949.00$						&$963.16$			& $972.48$				&$988.55$				& $-$ \\
$\Delta\chi^2$				& $-$							&$13.2$				& $23.5$					&$39.8$					& $-$ \\
$\chi^2_{\mathrm{UTas}}$	& $23.79$ 						&$24.83$				& $26.41$ 				&$28.86$					& $-$ \\
$\chi^2_{\mathrm{Danish}}$ 	& $79.77$							&$79.60$				& $80.75$ 				&$88.93$					& $-$ \\
$\chi^2_{\mathrm{OGLE}}$ 	& $845.50$ 						&$858.77$			& $865.24$ 				&$870.55$				& $-$ \\
\hline
$\tz$ 	  				& $4587.18  \pm   0.80$	 	 		&$4332.27\pm0.29$		& $4332.10\pm0.27$ 		&$4334.99\pm0.28$			& MHJD\\
$\te$ 	  				& $213.82  \pm   1.04$ 		 		&$52.00\pm3.63$		& $38.32\pm2.60$ 			&$53.46\pm0.81$			& days\\
$\alpha$ 	 				& $2.810  \pm    0.006$ 		 		&$3.227\pm0.030$		& $3.305\pm0.037$ 			&$4.570\pm0.018$			& rad\\
$\uz$ 					& $-1.573 \pm 0.013$ 				&$0.091\pm0.005$		& $0.164\pm0.019$ 			&$0.277\pm0.010$			& $-$\\
$\rho_{*}/10^{-3}$ 			& $0.34 \pm 0.01$ 					&$0.98\pm0.19$		& $1.55\pm0.16$ 			&$1.33\pm0.05$			& $-$\\
$d$ 		 				& $0.427 \pm 0.002$ 		 		&$0.673\pm0.011$		& $0.760\pm0.015$ 			&$2.158\pm0.0169$			& $-$\\
$q$ 		 				& $0.078 \pm 0.001$ 				&$0.177\pm0.017$		& $0.236\pm0.024$ 			&$0.288\pm 0.0096$		& $-$\\
$g(I)=\fb(I)/\fs(I)$			& $7.15\pm0.013$ 			             	&$68.11\pm0.013$             & $40.13\pm0.09$   	    		&$56.98 \pm 0.019$               	& $-$\\
$I_{s}$					& $17.89 \pm 0.01$ 				  	&$20.21 \pm 0.01$	       	 & $19.65\pm0.09$         	 	&$20.02 \pm 0.01$               	& $-$\\
$I_{b}$					& $15.75 \pm 0.01$ 				  	&$15.63 \pm 0.01$	       	 & $15.64\pm0.09$         	 	&$15.63 \pm 0.01$               	& $-$\\
$(V-I)_{s}$				& $1.80 \pm 0.10$ 				  	&$1.93 \pm 0.11$	       	 & $1.91\pm0.11$         	 	&$1.92 \pm 0.12$               	& $-$\\
$\theta_*$					& $1.18 \pm 0.24$ 				  	&$0.46 \pm 0.09$	       	 & $0.59\pm0.12$         	 	&$0.50 \pm 0.10$               	& $\mu as$\\
\hline \hline
Lens in the Disk			&&&&& \\
\hline
$M_1$ 			& $1.50_{-0.58}^{+1.85}$				& $0.42_{-0.22}^{+0.40}$ 		& $0.34_{-0.18}^{+0.36}$ 		& $0.34_{-0.18}^{+0.37}$ 		& $\msun$	\\ \\
$M_2$ 			& $0.12_{-0.05}^{+0.14}$				& $0.07_{-0.04}^{+0.07}$		& $0.08_{-0.04}^{+0.08}$		& $0.10_{-0.05}^{+0.11}$		& $\msun$	\\ \\
$\dl$ 			& $1.00_{-0.36}^{+0.95}$				& $ 5.7_{-1.5}^{+1.1}$ 		& $6.1_{-1.5}^{+1.1}$	 	& $6.1_{-1.5}^{+1.0}$		& $\kpc$		\\ \\
$v$ 				& $25_{-9}^{+24}$	 				& $80_{-21}^{+15}$	 		& $93_{-22}^{+16}$	 		& $67_{-16}^{+11}$			& $\mathrm{km\,s^{-1}}$	\\ \\
\hline \hline
Lens in the Bulge			&&&&& \\
\hline
$M_1$ 			& $41_{-14}^{+14}$					& $1.25_{-0.59}^{+1.47}$ 		& $0.79_{-0.35}^{+0.93}$ 		& $0.79_{-0.36}^{+0.94}$		& $\msun$	\\ \\
$M_2$ 			& $3.2_{-1.1}^{+1.1}$				& $0.22_{-0.10}^{+0.26}$		& $0.19_{-0.08}^{+0.22}$		& $0.23_{-0.10}^{+0.27}$		& $\msun$	\\ \\
$\dl$ 			& $6.7_{-0.7}^{+0.4}$				& $7.3_{-0.8}^{+0.6}$ 		& $7.3_{-0.8}^{+0.6}$	 	& $7.3_{-0.8}^{+0.6}$		& $\kpc$		\\ \\
$v$ 				& $167_{-17}^{+10}$ 				& $102_{-12}^{+8}$	 		& $111_{-12}^{+10}$		& $79_{-8}^{+6}$			& $\mathrm{km\,s^{-1}}$	\\ \\
\hline
  \end{tabular}
  \label{tab:partable}
\end{table*}

Now allowing for the parameters $d$ and $q$ to vary as well, we use our MCMC algorithm to find the
best solutions in each of these local minimum regions. These are identified with white filled circles on \Fig{fig:chi2mapC} and \ref{fig:chi2mapS} and correspond to the models listed in \Tab{tab:partable}, and shown in \Fig{fig:fit1}, 
\ref{fig:fit2}, \ref{fig:fit3} and \ref{fig:fit4}.
The best model lightcurve is dominated by strong caustics, which all viable
models must reproduce, with the low-magnification base PSPL curve barely noticeable. All models have
the first anomalous OGLE points on the descending side of the caustic entry except for the worst
model, model $W_c$, which has this OGLE point on the ascending part of the caustic
entry. Statistically, the former case is more likely to be observed since the ascending part of the
caustic entry happens much more rapidly than the descending side.

Our best model, $C_s$, has $\chi^2 = 949$ for 975 data points, with the other competitive models at
$\Delta\chi^2=13.2$ (model $C_c$), $\Delta\chi^2=23.5$ (model $I$)  and $\Delta\chi^2=39.6$ (model
$W_c$).

\begin{figure*}
  \centering
  \includegraphics[width=10.5cm, angle=270]{fig/modelCs.ps}
  \caption{Best-fitting binary lens model $C_s$ with residuals and a zoom on the anomaly (left
    inset). Data points are plotted with 1-$\sigma$ error bars. The trajectory of the source in the
    lens plane with the caustics is plotted as an inset in the top right corner of the figure, with
    the primary lens component located at the coordinate system's origin.  \label{fig:fit1}}
 
\end{figure*}

\begin{figure*}
  \centering
  \includegraphics[width=10.5cm, angle=270]{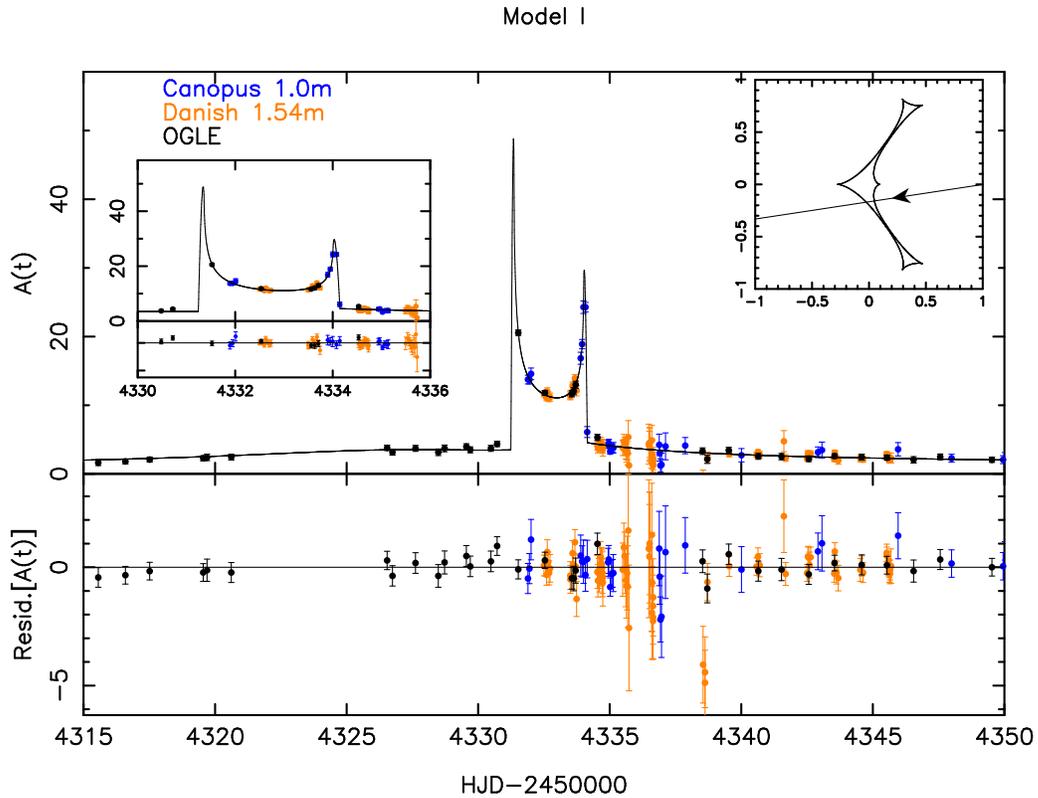}
  \caption{Same as \Fig{fig:fit1} for model $I$.   \label{fig:fit2}}
\end{figure*}

\begin{figure*}
  \centering
  \includegraphics[width=10.5cm, angle=270]{fig/modelCc.ps}
  \caption{Same as \Fig{fig:fit1} for model $C_c$.  \label{fig:fit3}}
\end{figure*}
\begin{figure*}
  \centering
  \includegraphics[width=10.5cm, angle=270]{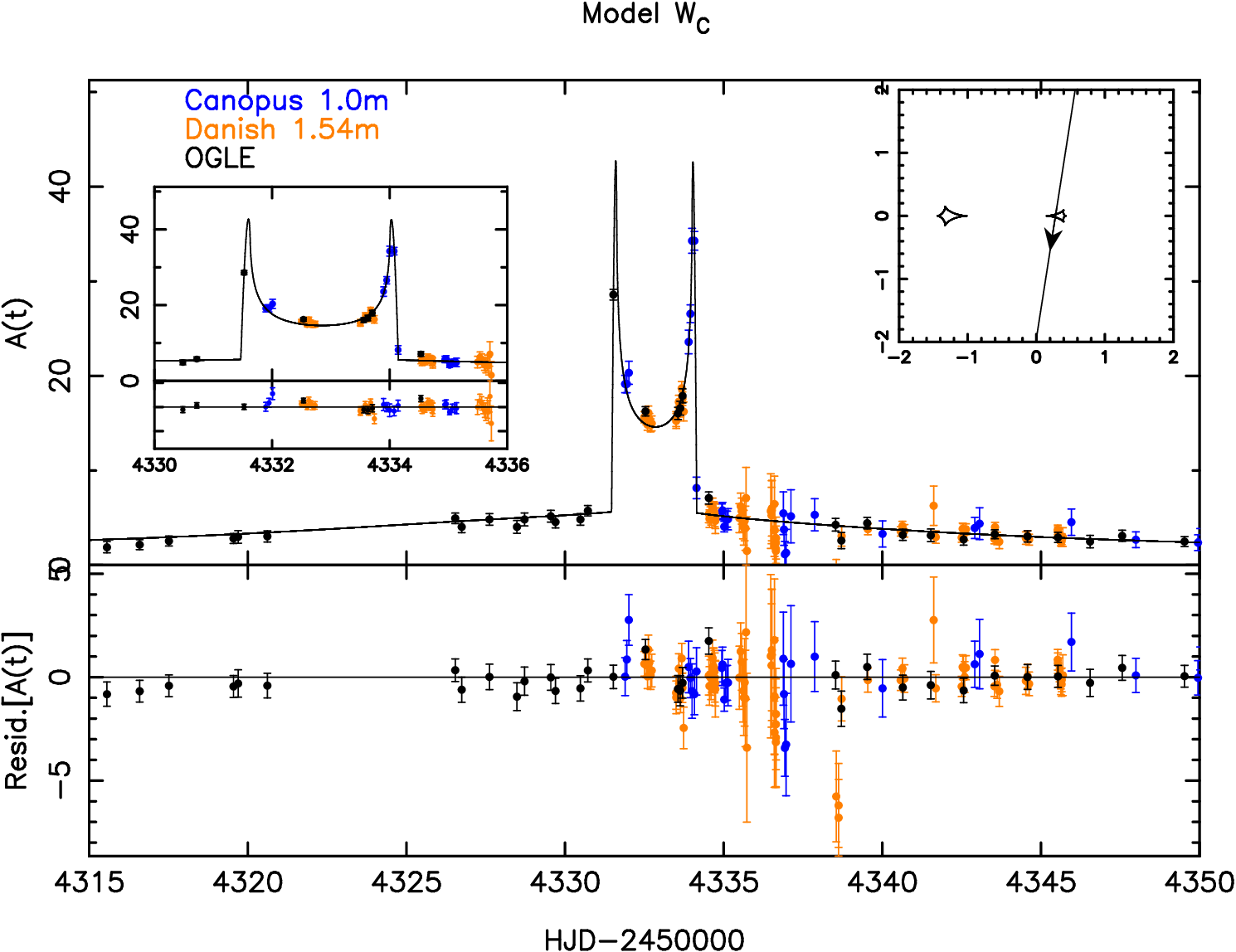}
  \caption{Same as \Fig{fig:fit1} for model $W_c$. \label{fig:fit4}}
\end{figure*}

\subsection{Discussion}

Fig. \ref{fig:tEmap} shows that the models with a source crossing a secondary caustic have increasingly large values of $\te$ as they go towards lower values of the mass ratio. This is
expected since the time $\Delta t$ between $\tentry$ and $\texit$ is fixed by the data. As the size of caustics scales with $q^{1/2}$, and $\te \sim \Delta t/q^{1/2}$,
the source must therefore cross the Einstein Ring over a longer timescale in order to conserve the right timing for $\sentry$ and $\sexit$. In addition to this,
blending decreases for decreasing values of $q$ , and therefore decreases with increasing $\te$, contrary to what might be expected. Indeed, one would expect the blending factor $g=\fb/\fs$ (where $\fb$ and $\fs$ are the blend and source flux respectively) to increase with increasing $\te$ in order to mask long timescales and reproduce the observed timescale. However in this region of parameter space, the caustics are weak, which means that too much blending would not allow models 
to reproduce the observed rise in the source magnitude at the caustic entry and caustic exit. For a region of parameter space to contain satisfactory models, there must therefore be a fine balance between blending, timescale and mass ratio.

For models where the source crosses a central caustic, the impact parameter $\uz$ must decrease with decreasing mass ratio, since the size of central caustic decreases with decreasing mass ratio, and the range of allowed $\uz$ decreases if the source must cross the caustic. This means that for smaller mass ratios, blending will have to increase in order to mask the correspondingly higher PSPL magnification of the source that results from the smaller impact parameter.

\subsection{Physical properties of the models} \label{sec:properties}

\subsubsection{Source characteristics}

\begin{figure}
	\centering
	\includegraphics[width=6cm, angle=270]{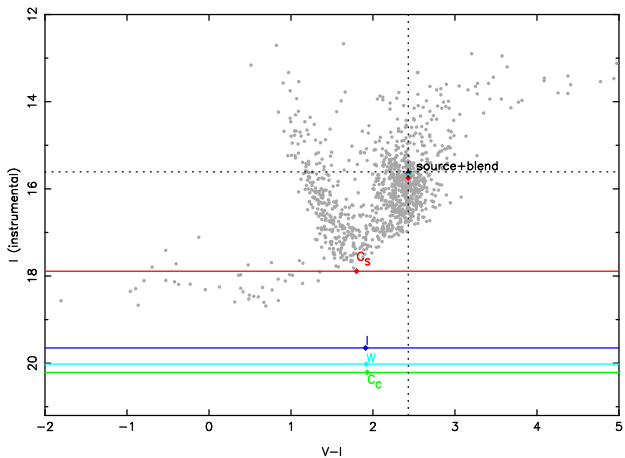}
	\caption{Colour - Magnitude diagram of the field. The target OGLE-2007-BLG-472 is shown as a black triangle at $(V-I, I)=(2.43, 15.61)$. The position of the deblended source for each model is labeled and indicated by a dotted line and a coloured diamond, with the blend for each model also plotted as a diamond in the same colour.}
	\label{fig:472CMD}
\end{figure}

A colour-magnitude diagram of the field (Fig. \ref{fig:472CMD}) was produced extracting 1497 stars
from I and V images at $t =4340.08$ (I) and $t=4340.13$ (V) taken at the Danish 1.54m telescope. The
combination of the source and the blend lies very slightly blueward of the red giant clump, at
$(V-I)$=2.43. All the models, however, are heavily blended (Table \ref{tab:partable}). The actual
source magnitude and blending magnitude for each model can be found using the equations
$I_s=I_{\mathrm{base}}+2.5\, \mathrm{log}(1+g)$ and $I_b=I_s-2.5\, \mathrm{log}(g)$.

Using this equation, we find source magnitudes ranging from 17.89 (model $C_s$) to 20.21 (model $C_c$) (see \Tab{tab:partable}). Our V-band data set does
not allow us to determine the source's colour, but assuming that the source is a main sequence star
we use the calculated I magnitude of the source for each model to estimate a colour, using the
results of \cite{holtzman98bulge}. This then enables us to estimate the source's angular radius
which we use in Section \ref{sec:lensprop} to compute probability densities of the lensing
system's properties.

We calibrate the baseline magnitude of our target (source and blend combined) using the location of the red clump as a reference. We find $I_{\mathrm{base}}=15.61 \pm 0.10$, which is in agreement with the OGLE value of  $I_{\mathrm{base}}=16.00 \pm 0.50$. Comparing this to the location of the red clump, we can derive an estimate for the reddening coefficient $\ai$. From Hipparcos results, \cite{stanekgarnavich98hipparcos} find an absolute magnitude for the red clump at $M_{\mathrm{I}, RC}=-0.23 \pm 0.03$. Using a distance modulus to the galactic centre of $\mu =14.41 \pm 0.09$ (i.e. assuming $\ds=7.6$ kpc) \citep{eisenhauer05sinfoni}, this translates to a dereddened magnitude for this target of $I_{\mathrm{base}} =14.18 \pm 0.09$. Hence using the relation $\ai=I_{\mathrm{base}}-M_{\mathrm{I}, RC}-\mu$, we get a value for the $I$-band reddening parameter of $\ai=1.43 \pm 0.13$. Alternatively, fitting 2MASS isochrones to our CMD, we obtain a value $\ai=1.46 \pm 0.08$ and $E(V-I)=1.46 \pm 0.11$. We use these values of reddening to determine dereddened magnitudes and colours for the source of each model. These, together with the surface brightness relations from \cite{kervella08}, allow us to calculate the apparent angular radius of the source $\theta_*$ for each of the models, given in \Tab{tab:partable}.

\subsubsection{Lens characteristics}\label{sec:lensprop}

Although the characteristics of any microlensing event depend on various properties of the lensing system, including the mass of the lenses, the only measurable quantity that can be directly related to physical properties of the lens is the timescale of the event $t_{\mathrm{E}}$.  While the physical properties of the lensing system can be fully constrained when the photometry is affected by both finite source-size effects and parallax, when these are not measured, such as is the case with our analysis OGLE-2007-BLG-472, we can still use Bayesian inference to determine probability densities of physical properties of the lens, based on a chosen Galactic model. We have chosen not to include parallax in our analysis because its effect would be very small for such a low-magnification event; in addition to this, we are only seeking a first-order analysis of binary-lens events with our current method, although second-order effects such as parallax and lens rotation will be taken into account in future work.

\begin{figure*}
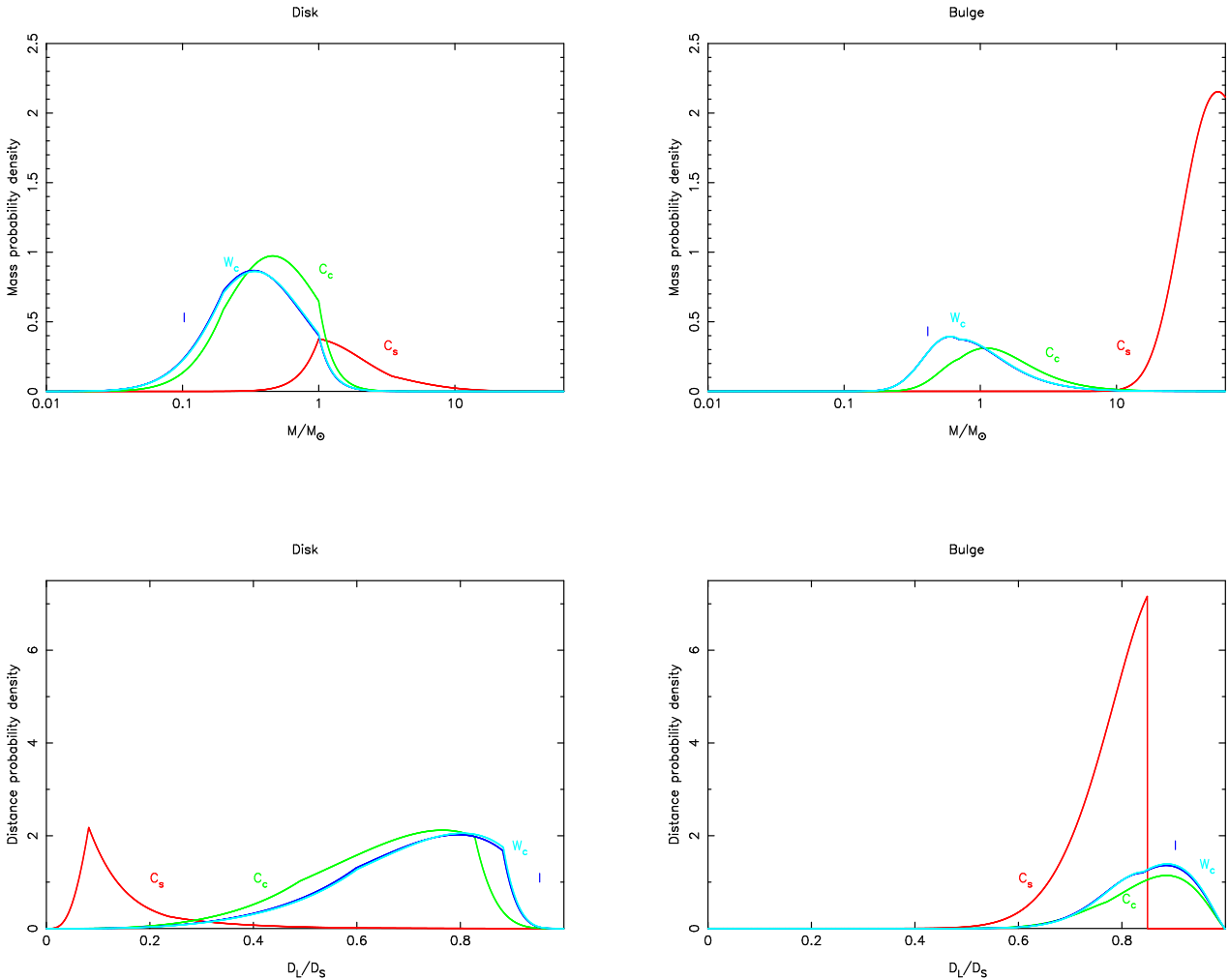

\centering
\subfigure
{
    \label{fig:sub:a}
    \includegraphics[width=6cm, angle=270]{fig/pmplot-disk.ps}
}
\hspace{1cm}
\subfigure
{
    \label{fig:sub:b}
    \includegraphics[width=6cm, angle=270]{fig/pmplot-bulge.ps}
}

\vspace{1cm}

\subfigure
{
    \label{fig:sub:c}
    \includegraphics[width=6cm, angle=270]{fig/pxplot-disk.ps}
}
\hspace{1cm}
\subfigure
{
    \label{fig:sub:d}
    \includegraphics[width=6cm, angle=270]{fig/pxplot-bulge.ps}
}

\caption{Probability densities for the mass of the primary lens star and the fractional distance $\dl/\ds$, for a lens in the disk (left side) and a lens in the bulge (right side). The values quoted in Tables \ref{tab:partable} \& \ref{tab:partable} are the median value and the limits of the 68.3\% confidence interval. On each plot, the probability densities are plotted for model $C_s$ (red), model $C_c$ (green), model $I$ (dark blue), and model $W_c$ (light blue).\label{fig:probabilities}}
\end{figure*}

\begin{figure}
	\centering
	\includegraphics[width=6cm, angle=270]{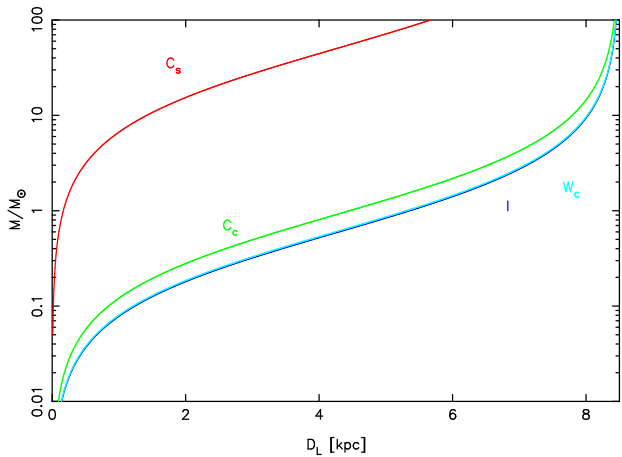}
	\caption{Mass-distance diagram showing the constraint on the lens mass from the source size, given by \Eq{eq:mx}, for each model. The curves are labeled with the name of the model to which they correspond. \label{fig:mass-distance}}
\end{figure}

We use our fitted value of the source size parameter $\rho_*$ to place constraints on the mass of the lens, which can be expressed as a function of fractional distance $x=\dl/\ds$ and the source size $\rho_*$ as (e.g. \citealt{dominik98})

\begin{equation}
\label{eq:mx}
\frac{M(x)}{\Msun}=\frac{c^2}{4G\Msun}\frac{\ds \,\theta_{*}^2}{\rho_{*}^2}\frac{x}{1-x}\, ,
\end{equation}
where $M$ is the mass of the lens, $\theta_*$ is the angular radius of the source, the value of which is given in \Tab{tab:partable}, and other quantities are defined as before. The mass-distance curve showing constraints from this equation is plotted on \Fig{fig:mass-distance}.

However, since we cannot measure parallax for this event, we use a probabilistic approach following that of \cite{dominik06stoch} to derive probability densities for physical properties of lens components. The Galactic model used here is a piecewise mass spectrum (e.g. \citealt{chabrier03}), two double exponentials for the disk mass density and a barred bulge tilted at an angle of $20^{\circ}$ with the direction to the Galactic centre (\citealt{dwek95}), and the distribution of effective transverse velocities used in \cite{dominik06stoch}.

Using these galactic models, we infer properties for the lensing system, separating the cases where the lens is in the Galactic disk and in the Galactic bulge. For a lens in the disk, we find a primary mass $1.50_{-0.58}^{+1.85} \Msun$ and a secondary mass of  $0.12_{-0.05}^{+0.14} \Msun$, at a distance of $1.00_{-0.36}^{+0.95}$ kpc with a lens velocity of $25_{-9}^{+24}\, \mathrm{km\,s^{-1}}$. For a lens in the bulge, we find a primary mass $41_{-14}^{+14} \Msun$ and a secondary mass of  $3.2_{-1.1}^{+1.1} \Msun$, at a distance of $6.7_{-0.7}^{+0.4}$ kpc with a lens velocity $167_{-17}^{+10}\, \mathrm{km\,s^{-1}}$. These are the physical lens properties for the lowest-$\chi^2$ model (model $C_s$). The values of these physical parameters for the other models are given in Table \ref{tab:partable}. Probabilty densities of these properties for all models are plotted on Fig. \ref{fig:probabilities}. 

\subsubsection{Discussion}

For our lowest-$\chi^2$ model, the parameters we find imply very unusual properties of the lensing system. As discussed in \Sec{sec:excluding}, the fact that we find these types of models is a consequence of the fitting approach we are taking. Traditional fitting methods would struggle to find these minima, since most of them require providing a starting point in parameter space. This is an issue when solely using an MCMC algorithm to fit microlensing events: although an MCMC run may be able to make its way through parameter space to find minima reasonably far away from its starting point, it is highly unlikely that a chain will be able to reach a minimum that has parameters different from the starting point by more than one order of magnitude. As we see from Fig. \ref{fig:tEmap}, there exist minima in many parts of parameter space, with values of $\te$ that are different by almost two orders of magnitude. These parameters are non-intuitive, since they cannot be guessed only by looking at the lightcurve. As a result, it is improbable that this kind of parameters will be used as starting points for "classic" fitting algorithms. 

We solve this problem for the static binary-lens case by resorting to the method described in \Sec{sec:exploration}. Using this approach, we manage to systematically locate minima in parameter space. However, one then has to be careful with interpreting the significance of the obtained model parameters. The shape of probability densities shown in \Fig{fig:probabilities} for model $C_s$ indicates that our value of $\te$ push the lens mass towards the end of the adopted mass spectrum in the Galactic model we have adopted. This results in the abrupt transients seen on \Fig{fig:probabilities}. Similarly, the mass-distance curve for model $C_s$ on \Fig{fig:mass-distance} shows that the mass of the lens increases very rapidly for lenses above $\sim 1$ kpc. These unusual curves are caused by a value of $\te \sim 200$ days. Models with $\te \sim 3000$ days (corresponding to the low-$q$ minimum visible on \Fig{fig:chi2mapS} \& \ref{fig:tEmap}) are obviously not acceptable, but how can we formally reject them? Finding these models from minima in the $\chi^2$ surface shows the limits of using $\chi^2$ as a strong criterion for favouring models. A solution to this would be to use prior distributions on as many of the parameters as we can. During the MCMC part of our fitting process, this would mean that we obtain posterior distributions that are different from the ones obtained without using prior distributions on the parameters, or, equivalently, assuming uniform priors for all parameters. Such priors can be obtained in various ways, such as looking at the distribution of timescales for past microlensing events or calculating these distributions from Galactic models (e.g. \citealt{dominik06stoch}), or by using luminosity functions of the Galactic bulge to find a prior for the blending factor $g$ (e.g. \citealt{holtzman98bulge}). Such work requires careful consideration of which priors are most appropriate to use, and is beyond the scope of this paper. Using these priors in combination with our method to find minima will lead to more robust determination of minima by taking into account our knowledge of physical parameter distributions.

\section{Summary and prospects}

Our analysis of OGLE-2007-BLG-472 is a good illustration of the importance and power of using parameters that are related to actual observed features. Indeed, despite incomplete coverage of the caustic entry and high blending, a few crucial data points and an appropriate choice of non-standard parameters enable us to find several good binary-lens model fits to our data for this event by exploring the parameter space systematically.  Some of the good fits that we identify have unphysical parameters, and we must then reject them. However using this parameterisation allows us to be certain that the parameter space has been thoroughly explored. We find four models with different parameters: two close binary models, one intermediate configuration, and a wide binary model. The lowest-$\chi^2$ model corresponds to a G dwarf star being lensed by a binary system with component masses $1.50_{-0.58}^{+1.85} \Msun$ (for the primary) and $0.12_{-0.05}^{+0.14} \Msun$ (for the secondary), which are compatible with our blending values. However it is obvious from physical parameter distributions that using $\chi^2$ as a sole criterion for determining the best model is insufficient, because it does not take into account our knowledge of the distributions of physical parameters. 

Since the approach presented in this paper can form the basis for a systematic, wide ranging exploration of the parameter space to localise all possible models for a given data set, it is particularly relevant to current efforts to automatise real-time fitting of binary-lens events. This could prove useful to provide faster feedback on the events being observed and prioritise observing schedules, especially on robotic telescopes. Expanding robotic telescope networks controlled by automated intelligent algorithms are expected to play an increasingly important role in microlensing surveys in the coming years (e.g. \citealt{tsapras08}). Fitting methods such as the one described in this paper are essential for making sure any anomalies are interpreted correctly, and that minima are located in as large a part of parameter space as possible.

\section*{Acknowledgements}

NK acknowledges STFC studentship PA/S/S/2006/04497 and an STFC travel
grant covering his observing run at La Silla. We thank David Warren for financial support for the Mt Canopus Observatory. NK thanks Pascal
Fouqu\'{e} for organising a workshop in Toulouse in November 2007, and
Joachim Wambsganss and Arnaud Cassan for their invitation to visit the
Astronomisches Rechen-Institut in Heidelberg in April 2008. 
We would like to thank the anonymous referee for his helpful comments on
the manuscript. We also thank the University of Tasmania for access
to their TPAC supercomputer on which part of the calculations were
carried out. PF expresses his gratitude to ESO for a two months
invitation at Santiago headquarters, Chile in October and November
2008. The OGLE project is partially supported by the Polish MNiSW
grant N20303032/4275.  

\bibliographystyle{mn2e}
\bibliography{iau_journals,microbib_mnras}
\bsp

\newpage

\label{lastpage}

\end{document}